# Mechanical Properties of Penta-Graphene Nanotubes

Mingwei Chen[1,2], Haifei Zhan[2,3,*], Yinbo Zhu[1], Hengan Wu[1,*], and Yuantong Gu[2]

[1]CAS Key Laboratory of Mechanical Behavior and Design of Materials, Department of Modern Mechanics, CAS Center for Excellence in Nanoscience, University of Science and Technology of China, Hefei, Anhui 230027, China
[2]School of Chemistry, Physics and Mechanical Engineering, Queensland University of Technology (QUT), Brisbane, QLD 4001, Australia
[3]School of Computing, Engineering and Mathematics, Western Sydney University, Locked Bag 1797, Penrith NSW 2751, Australia

**ABSTRACT**

Penta-graphene is the name given to a novel puckered monolayer of carbon atoms tightly packed into an inerratic pentagonal network, theoretically, which exhibits excellent thermal and mechanical stability and can be rolled into penta-graphene nanotubes (PGNTs). Herein, we perform the first simulation study of mechanical properties of PGNTs under uniaxial tension. In addition to comparable mechanical properties with that of carbon nanotubes (CNT), it is found that PGNTs possess promising extensibility with typical plastic behavior due to the irreversible pentagon-to-polygon structural transformation and the hexagon carbon ring becomes the dominating structural motif after the transformation. The plastic characteristic of PGNTs is inherent with strain-rate and tube-diameter independences. Moreover, within ultimate temperature ($T < 1100$ K), tensile deformed PGNTs manifest similar phase transition with an approximate transition ratio from pentagon to hexagon. The intrinsic insight provides a fundamental understanding of mechanic properties of PGNTs, which should open up a novel perspective for the design of plastic carbon-based nanomaterials.

**INTRODUCTION**

Carbon has abundant and various allotropes in different dimensions, ranging from three-dimensional (3D) diamond and graphite to well-known two-dimensional (2D) graphene,[1] quasi-one-dimensional (1D) nanotubes,[2] and zero-dimensional (0D) fullerenes.[3] In the past few decades, extensive studies have been carried out to probe the excellent physical, electronic, and mechanical properties of low-dimensional carbon polymorphs and their derivatives, both in scientific research and engineering application. Triggered by the intriguing behavior of low-dimensional carbon nanomaterials, myriad of novel 2D carbon allotropes have been predicted theoretically and even been synthesized successfully (*e.g.*, graphdiyne). Recently, Zhang *et al.* proposed a new 2D metastable carbon polymorph, penta-graphene, composed entirely of carbon pentagons and resembling the Cairo pentagonal tiling, which can be exfoliated from T-12 carbon.[4] First-principle calculations demonstrated that penta-graphene possesses dynamical, thermal and mechanical stability.[4-5] Several followed studies found that penta-graphene exhibits lower thermal conductivity compared with graphene[6-7] and its nanoribbon structure has a relatively stable band gap.[5] Sun *et al.* characterized the mechanical behavior of monolayer penta-graphene under multi-axial loading using density functional theory (DFT) calculations which indicates that penta-graphene has lower Young's modulus compared to graphene (in the same order of magnitude).[8]

Penta-graphene monolayer can be rolled into a pentagon-based nanotube, namely, penta-graphene nanotube (PGNT). Theoretical calculations have confirmed the mechanical and thermal stability of PGNT and demonstrated that PGNTs are semiconducting with regardless of chirality.[4] As a new 1D carbon nanostructure, a natural question arises, how it may differ from existing 1D carbon nanomaterials, such as carbon nanotube (CNT) and the recently synthesized diamond nanothread.[9-10] From the perspective of morphology, although analogue to CNT, PGNT has rough inner and outer surfaces due to the corrugated nature of penta-graphene. CNT has been widely used as a reinforcement in some nanocomposites to make them self-lubricating due to their lubricious nature,[11-12] PGNT may also own a potential application prospect similar to CNT. To our knowledge, the fundamental understanding of mechanical properties of PGNT is still lacking. Earlier studies found that penta-graphene begins to degenerate to a graphene-like hexagonal structure at a critical strain of 5.1%, and behaves plastically beyond this critical strain.[13] Inspired by this, it is of great interest to probe whether the tube structure will inherit such unique feature.

Therefore, based on large-scale molecular dynamics (MD) simulations, this work is intended to probe the mechanical properties of the PGNT. From the uniaxial tensile test, we observe a clear transition from pentagonal form to hexagonal structure and a plastic behavior associated with this transition process, which does not change with the varying diameter or temperature.

**COMPUTATIONAL METHODS**

Using classical molecular dynamics (MD) simulations, mechanical properties of PGNTs under uniaxial tension are investigated in this paper. As illustrated in Figure 1, PGNT can be formed by rolling up a penta-graphene sheet along the ($n$, $m$) chiral vector. The radius ($R$) and unit length ($L_{unit}$) of the tube can be estimated from $R = a\sqrt{n^2 + m^2}/2\pi$ and $L_{unit} = a\sqrt{n^2 + m^2}$, respectively ($a$ = 3.64 Å is the lattice constant of penta-graphene).[4] The DFT calculations suggested that PGNT is stable when $m = n$. To assess the mechanical properties of PGNT, a representative (8,8) nanotube with the length of ~16 nm has been selected. The ReaxFF reactive force filed was adopted to describe the atomic interactions between carbon atoms, which is a general bond-order-dependent force field that provides accurate descriptions of bond breaking and bond formation.[14] This force field has been successfully applied to investigate the chemomechanical behavior of graphene under impact,[15] as well as the mechanical properties of graphyne,[16] carbine,[17] and diamond nanothread.[17]

The initial equilibrium structure of PGNT was achieved by conjugate gradient energy minimization method. Then the system was equilibrated using Nosé-Hoover thermostat[18-19] for 100 ps. Periodic boundary conditions were applied along the length direction during the relaxation process. A low temperature of 10 K was selected to alleviate the influence from thermal fluctuations. After relaxation, the structure was switched to non-periodic boundary conditions, and a constant velocity (2 × 10⁻⁵ nm/fs) was applied to the right end of the tube (with the other end being fixed). The sizes of two fixed ends were both set as ~3 nm, and thus the strain rate is about 2 × 10⁻⁶ fs⁻¹. The simulation was continued until the failure of the tube. A time step of 0.05 fs was used for the velocity-Verlet integrater. All MD simulations were performed using the LAMMPS program.[20]

During the uniaxial tension, the atomic virial stress was calculated as:[21]

$$p_i^{ab} = \frac{1}{\bar{\omega}_i}\left\{-m_i v_i^a v_i^b + \frac{1}{2}\sum_{j \neq i} F_{ij}^a r_{ij}^b\right\}, \qquad (1)$$

where $\bar{\omega}_i$, $m_i$, and $v_i$ denote the effective volume, mass, and velocity of atom $i$, respectively. $F_{ij}$ and $r_{ij}$ are the force and distance between atoms $i$ and $j$; and the indices $\alpha$ and $\beta$ represent the Cartesian components. To estimate the effective atomic volume, PGNT is approximately regarded as a hollow cylinder with a thickness of the penta-graphene (~3.68 Å). The thickness of the penta-graphene is estimated by simulating a multiple layer penta-graphene structure with periodic boundary conditions in the thickness direction. The virial stress of PGNT can thus be obtained by averaging the atomic stress.

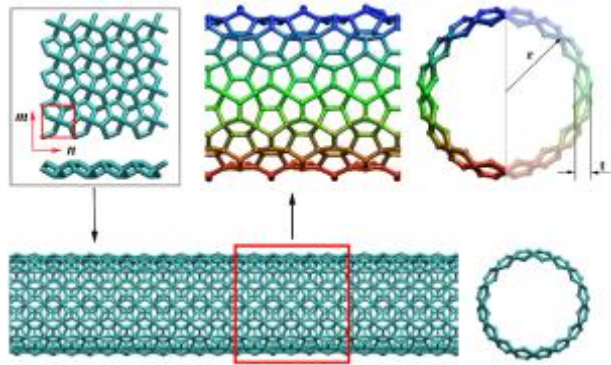

**Figure 1.** Atomic structure of PGNT (8,8). The top-left shows the structure of penta-graphene and its unit cell. The top-right shows the partial enlarged detail of PGNT.

**RESULTS AND DISCUSSION**

*Tensile behavior.* Figure 2a illustrates the stress-strain curve of the (8,8) PGNT under uniaxial tension. Indeed, there exists an evident plastic deformation period besides the elastic portion, similar as observed from penta-graphene.[13] In first, PGNT behaves elastically until the strain approaches 0.062 (with the stress around 56 GPa). The inflection point is marked as "A" to denote the end of the elastic stage. Then, PGNT manifests obvious plastic behavior with large strain range ($0.062 < \varepsilon < 0.61$) and stress fluctuation ($25$ GPa $< \sigma < 100$ GPa). In particular, in this plastic deformation, the stress-strain curve first increases and then suddenly drops by about 60 GPa at $\varepsilon = 0.12$. Subsequently, the stress displays a jagged increase with further tensile loading. When the strain is beyond 0.61, the stress suddenly drops to 0 GPa (marked point "B"), indicating the final failure of PGNT. Moreover, the elastic and plastic deformations are also well demonstrated by the changes of potential energy (PE) density shown in Figure 2b. The inset in Figure 2b indicates a good parabolic relationship between the PE density and the tensile strain in the elastic stage. After the inflection point, the PE density exhibits three different regions, corresponding to different features in the plastic deformation.

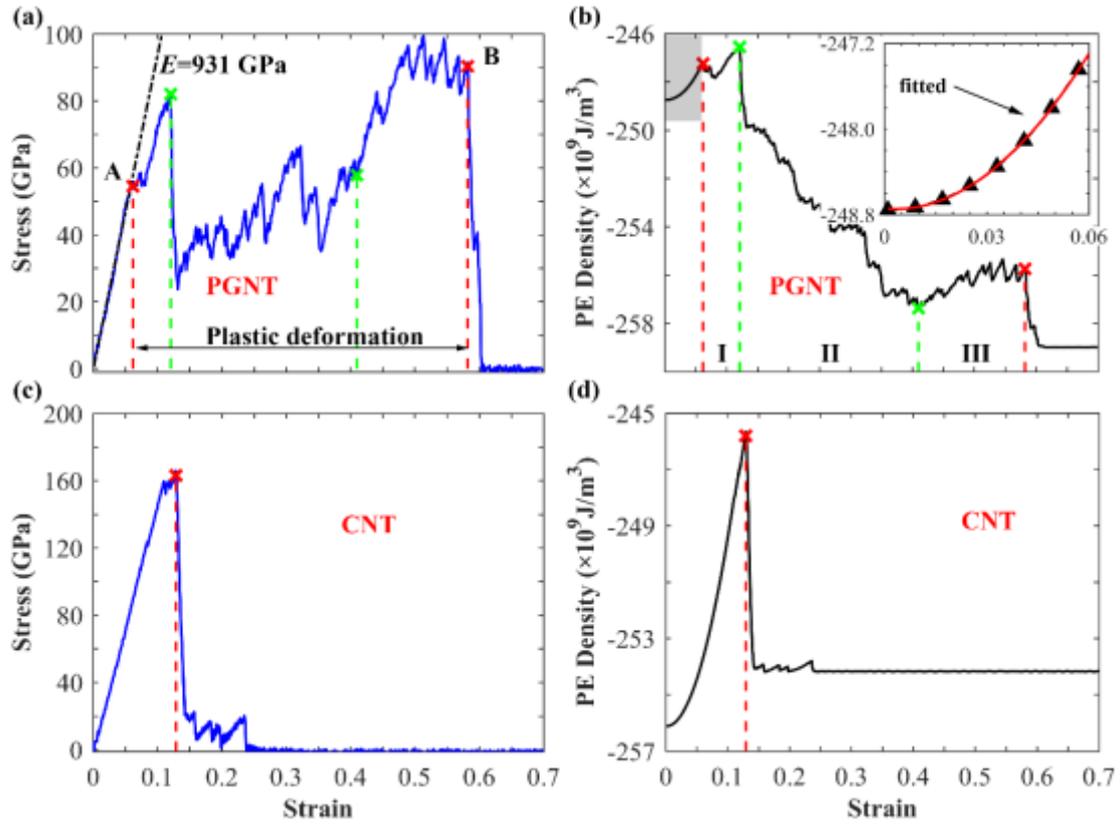

**Figure 2.** Comparisons of the mechanical properties between (8,8) PGNT and (10,10) CNT. (a) and (b) The stress and PE density as functions of tensile strain in the simulation of PGNT. The inset in (b) shows the fitting of PE density in the elastic stage. (c) and (d) The stress-strain curve and corresponding PE density of CNT.

To assess the mechanical properties, the linear regime of stress-strain curve in the elastic stage is fitted through $\sigma = E(\varepsilon - \varepsilon_0) + \sigma_0$ using the linear regression,[17, 22-24] which yields to an effective Young's modulus $E$ = ~931 GPa. This value is closed to that extracted from the PE curve according to $U = E\varepsilon^2/2 + U_0$ (about 895 GPa). According to Figure 2a, the (8,8) PGNT has a yield strength of 56 GPa and a failure strength of 90 GPa. Corresponding yield strain and failure strain are 0.062 and 0.61, respectively. Of interest, we carried out a simulation of (10,10) CNT with the same simulation setup as a comparison, which possesses a similar diameter with that of (8,8) PGNT. In contrast, CNT exhibits a classical brittle characteristic without plastic deformation (Figure 2c,d). Although CNT has higher elastic modulus (~1477 GPa) and yield strength (~160 GPa), the failure strain of CNT is only ~0.15, which is much smaller than that of PGNT.

***Deformation mechanisms.*** To gain more insight into the deformation mechanisms of PGNT under uniaxial tension, atomic configurations of PGNT during the tensile process are investigated. For such purpose, the number of different carbon polygons is counted (Figure 3a), including trigonal (C3), tetragonal (C4), pentagonal (C5), hexagonal (C6), heptagonal (C7), and octagonal (C8). As expected, during the elastic deformation period, only bond stretching event occurs, and thus all of carbon polygons are C5. The atomic configuration in Figure 3b indicates that there is a strong local stress concentration in the structure when $\varepsilon = 0.06$.

When the strain is beyond the yield strain, the deformation displays typical plastic behavior. The plastic deformation stage of PGNT can be divided into three sub-stages. In the stage-I, the number of C5 shows a sharp drop due to the bond breaking among adjacent pentagons (Figure 3c), whereas, the number of C8 increases significantly. The stress-strain curve in Figure 2a exhibits a transient jagged platform followed by a linear-like increase in this stage, demonstrating the bond breaking of C5 and the formation of more defects. In the stage-II, the number of different polygons exhibits remarkable changes. Note that, a sudden drop of stress-strain curve arises at the boundary between the stage-I and stage-II. Correspondingly, numbers of C5 and C8 display sudden inverse changes. Bond breaking at the end of stage-I has greatly released the local stress. The rapid drop of C8 and the dramatic change of C5 at $\varepsilon = 0.12$ indicate the bond rebuilding of dangling atoms near defects. With increasing tensile strain, the number of C5 shows a gradual decrease, while that of C6 displays a continuous increase from zero and reaches a stable value (~400) when $\varepsilon > 0.4$. Those significant changes of numbers of different polygons signify a phase transition phenomenon in this sub-stage. Meanwhile, the corresponding stress-strain curve manifests drastic jagged fluctuation. In the beginning of stage-III, the number of each polygon reaches corresponding stable value, indicating the formation of metastable structure and the finish of pentagon-to-polygon transformation. In particular, C6 becomes the dominating structural feature due to its large proportion (> 50%). Thus, the deformed PGNT can be regarded as a CNT with some geometrical defects (*e.g.*, C5 and C7). Note also that, the stress-strain curve (Figure 2a) in this sub-stage shows a linear-like sharp increase from 50 GPa to 100 GPa with the strain ranging from 0.41 to 0.5, which is mainly due to that bonds are being stretched again without transformation of polygonal carbon rings. The stable tendency of the number of each polygon has continued until the failure of the whole structure (Figure 3e).

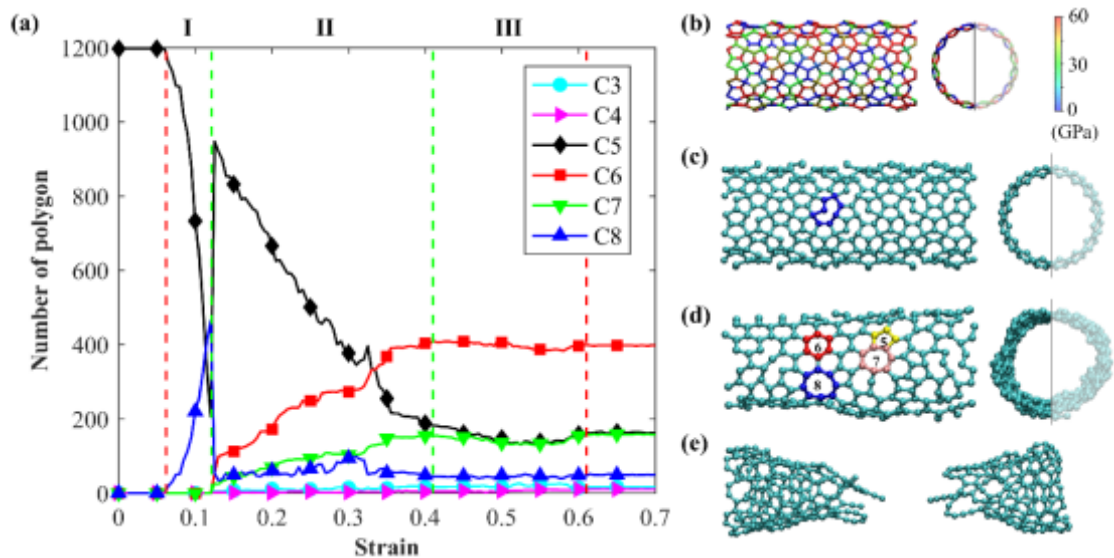

**Figure 3.** (a) Statistical numbers of different polygons as functions of the tensile strain. Atomic configuration of the PGNT: (b) $\varepsilon = 0.06$, colored by atomic stress, (c) $\varepsilon = 0.1$, stage-I of plastic deformation, (d) $\varepsilon = 0.3$, stage-II of the plastic deformation, and (e) $\varepsilon = 0.6$, the failure of PGNT. Only half of the tube is visible in (b-d) for visualization clearance.

To further verify the inherent plastic deformation of PGNT, we examined the structure under different strain rates and loading modes. We firstly verified the elasticity of PGNT, that is, the strained PGNT can fully recover to its initial structure in the release process when the strain is lower than the elastic strain limit (~0.061). Whereas, when the strain $\varepsilon > 0.061$, the deformed PGNT cannot recover due to the irreversible structural transformation. These results further approve that PGNT undergoes a plastic deformation during the tensile process. More importantly, it is worth noting that the strain rate might influence the deformation behavior of PGNT. In this regard, we chose another two different strain rates ($1 \times 10^{-6}$ fs$^{-1}$ and $4 \times 10^{-6}$ fs$^{-1}$), which yield to almost overlapped stress-strain curves (Figure 4) with Young's modulus of 895 and 886 GPa (calculated by PE density fitting), respectively. Therefore, the strain rate ($2 \times 10^{-6}$ fs$^{-1}$) applied in this study exerts ignorable impacts on the deformation behavior. We also employed another tensile loading scheme, *i.e.*, imposing a constant strain rate to the whole structure rather than a constant velocity to one end. Again, we obtained similar simulation results with clear elastic and plastic deformations (magenta line with rhombic symbols in Figure 4). Overall, unlike CNT, PGNT exhibits inherited plasticity from the penta-graphene, and experiences a significant irreversible structural phase transition during the plastic deformation.

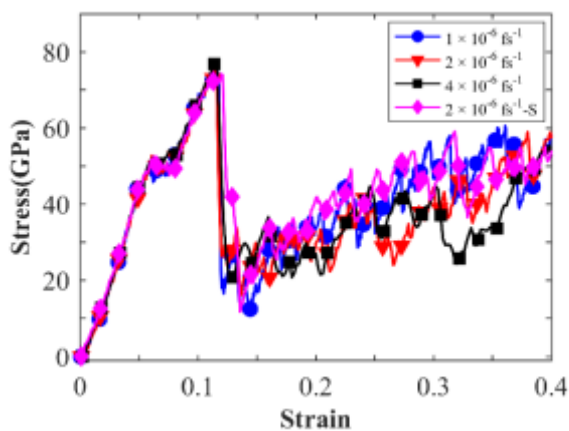

**Figure 4.** Stress-strain curves of the (8,8) PGNT under different strain rates. The "S" represents the PGNT being deformed by applying a constant strain rate.

***Diameter and temperature impacts.*** We further assessed the generality of the plastic characteristic of PGNT. Initially, we simulated PGNTs with different ($n$, $m$) indexes ranging from (4,4) to (12,12). It is found that all PGNTs exhibit a similar deformation behavior, *i.e.*, an evident plastic deformation period with phase transition after the elastic deformation. As compared in Table 1, the effective Young's modulus is generally independent of the diameter, which is same as observed in CNT.[25] Meanwhile, no clear relationship is observed between the diameter and yield strength/strain or the failure strength/strain. Strikingly, comparing the yield strain and the failure strain, all examined PGNTs manifest a similar plastic strain range (~0.06 < $\varepsilon$ < ~0.6) where the plastic deformation occurred.

**Table 1.** Mechanical properties of PGNTs with different diameters. $E$ is Young's modulus (GPa), $\sigma_y$ and $\sigma_f$ represent yield strength (GPa) and failure strength (GPa); $\varepsilon_y$ and $\varepsilon_f$ represent yield strain and failure strain. Note that, for some cases monoatomic chain will be generated at the deformation front

of the PGNT before final failure. In this scenario, the strain/stress before the formation of monoatomic chain is regarded as the failure strain/strength.

| (m, n) | (4,4) | (5,5) | (6,6) | (7,7) | (8,8) | (9,9) | (10,10) | (11,11) | (12,12) |
|---|---|---|---|---|---|---|---|---|---|
| $E$ | 704 | 721 | 833 | 880 | 895 | 899 | 895 | 888 | 902 |
| $\sigma_y$ | 38.4 | 49.0 | 53.8 | 30.2 | 55.9 | 57.9 | 46.9 | 49.3 | 38.6 |
| $\sigma_f$ | 83.9 | 100.2 | 98.42 | 93.85 | 90.3 | 93.6 | 96.9 | 91.1 | 84.3 |
| $\varepsilon_y$ | 5.1 | 7.8 | 5.9 | 5.9 | 6.2 | 6.1 | 6.3 | 6.0 | 6.3 |
| $\varepsilon_f$ | 42.7 | 60.8 | 55.4 | 55.2 | 58.2 | 53.8 | 52.1 | 55.0 | 60.7 |

The potential temperature influence on mechanical properties of PGNTs is also acquired. Specifically, we compared the mechanical performance of the (8,8) PGNT under the temperature ranging from 10 to 2000 K. It is found that at higher temperature (> 1100 K), PGNTs will undergo phase transitions during relaxation processes (without any external loading). Such thermal-induced phase transition phenomenon is consistent with the previous studies on penta-graphene, in which high kinetic energy (*i.e.*, temperature) is found to overcome the energy barrier and trigger the transition from pentagonal to hexagonal carbon rings.[13] The observation also agrees with the *ab initio* calculations, in which penta-graphene is found to be able to withstand a temperature about 1000 K.[4]

For other examined temperatures from 10 to 1100 K, similar phase transition processes were detected during the plastic deformation. As illustrated in Figure 5a, the carbon ring ratio shares a similar profile under different temperatures. Here, we propose the carbon ring ratio, $\mu^{CR} = N_{C6}/N_{C5}^0$, as an indicator of the phase transition, which is defined as the ratio between the number of hexagonal carbon rings $N_{C6}$ and the number of initial pentagons $N_{C5}^0$. According to Figure 5a, the carbon ring ratio maintains zero during the elastic deformation period. Thereafter, it increases with strain and saturates to a certain value until the final failure. Strikingly, the final carbon ring ratio $\mu_{ave}^{CR}$ is found to fluctuate around a value of 0.34 when temperatures change from 10 to 1100 K. Here, $\mu_{ave}^{CR}$ is estimated by averaging $\mu^{CR}$ after final fracture (i.e., between strain 0.7 to 0.9). We note that $\mu^{CR}$ maintains almost constant after failure with further simulation, indicating maximum phase transition during the tensile deformation. Such observation can be explained from the perspective of the maximum allowable transition ratio. That is, for a given PGNT, the total energy required to achieve the maximum transition ratio from pentagon to hexagon is a constant and independent of the contribution between tensile strain energy and kinetic energy. It is regarded that the accumulated tensile strain energy is able to lead to the full phase transition of PGNTs whatever the temperature is employed, and therefore induces a similar $\mu_{ave}^{CR}$ as observed in Figure 5b.

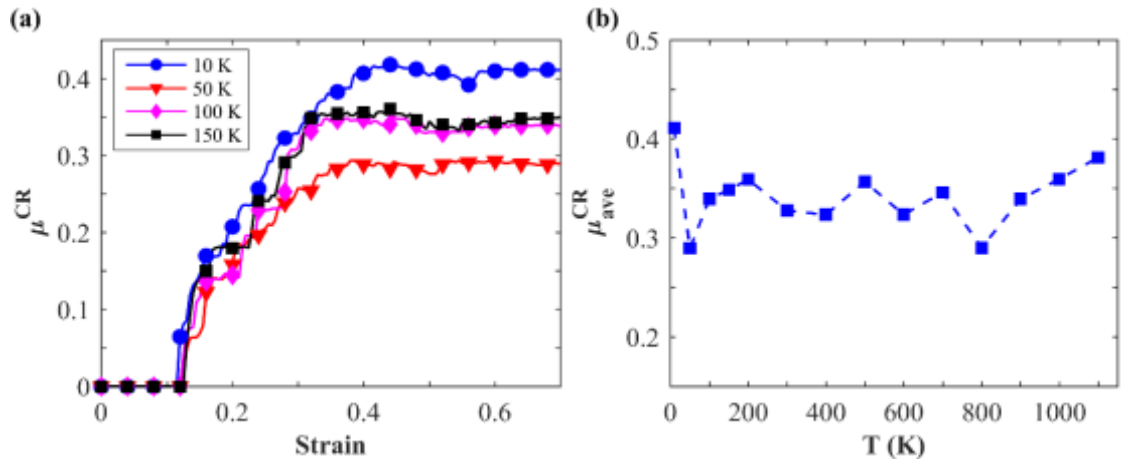

**Figure 5.** (a) The carbon ring ratio $\mu^{CR} = N_{C6}/N_{C5}^0$ as a function of strain at different temperatures for (8,8) PGNT. (b) The final carbon ring ratio $\mu_{ave}^{CR}$ as a function of temperature from 10 to 1100 K.

**CONCLUSION**

In summary, the mechanical properties of penta-graphene based nanotube structure (PGNT) have been acquired through large-scale MD simulations. It is found that PGNT has comparable mechanical properties with that of the CNT. However, unlike the brittle CNT, PGNT exhibits a large extensibility (with failure strain exceeding 60%) and behaves plastically during tensile deformation. The plasticity is inherently originated from the phase transition from pentagonal to polygonal (including trigonal, tetragonal, hexagonal, heptagonal, and octagonal) carbon rings. The plastic characteristic of PGNTs is intrinsic with tube-diameter and strain-rate independences. Additionally, within ultimate temperature ($T < 1100$ K), tensile deformed PGNTs manifest similar phase transition with an approximate transition ratio from pentagon to hexagon. For a given PGNT, the total energy required to achieve the maximum transition ratio from pentagon to hexagon is a constant and independent of the contribution between tensile strain energy and kinetic energy. This study has provided a fundamental understanding of the mechanical properties of PGNTs under uniaxial tension, which should shed lights on the design and application of plastic carbon-based nanostructures.

## AUTHOR INFORMATION

**Corresponding Authors**

*E-mail: h.zhan@westernsydney.edu.au

*E-mail: wuha@ustc.edu.cn

**Notes**

The authors declare no competing financial interests.

## ACKNOWLEDGEMENTs

This work was jointly supported by the National Natural Science Foundation of China (11525211), the Strategic Priority Research Program of the Chinese Academy of Sciences (XDB22040402), and Australian Research Council Discovery Grant DP170102861. The High Performance Computer resources provided by the Queensland University of Technology are gratefully acknowledged.

## REFERENCES

1. Geim, A. K., Graphene: Status and Prospects. *Science* **2009**, *324*, 1530-1534.
2. Iijima, S.; Ichihashi, T., Single-Shell Carbon Nanotubes of 1-nm Diameter. *Nature* **1993**, *363*, 603-605.
3. Kroto, H.; Heath, J.; O'Brien, S.; Curl, R.; Smalley, R., C60: Buckminsterfullerene. *Nature* **1985**, *318*, 162-163.
4. Zhang, S.; Zhou, J.; Wang, Q.; Chen, X.; Kawazoe, Y.; Jena, P., Penta-Graphene: A New Carbon Allotrope. *Proceedings of the National Academy of Sciences* **2015**, *112*, 2372-2377.
5. Rajbanshi, B.; Sarkar, S.; Mandal, B.; Sarkar, P., Energetic and Electronic Structure of Penta-Graphene Nanoribbons. *Carbon* **2016**, *100*, 118-125.
6. Xu, W.; Zhang, G.; Li, B., Thermal Conductivity of Penta-Graphene from Molecular Dynamics Study. *The Journal of chemical physics* **2015**, *143*, 154703.

7.  Wang, F. Q.; Yu, J.; Wang, Q.; Kawazoe, Y.; Jena, P., Lattice Thermal Conductivity of Penta-Graphene. *Carbon* **2016**, *105*, 424-429.

8.  Sun, H.; Mukherjee, S.; Singh, C. V., Mechanical Properties of Monolayer Penta-Graphene and Phagraphene: A First-Principles Study. *Physical Chemistry Chemical Physics* **2016**, *18*, 26736-26742.

9.  Fitzgibbons, T. C.; Guthrie, M.; Xu, E.-s.; Crespi, V. H.; Davidowski, S. K.; Cody, G. D.; Alem, N.; Badding, J. V., Benzene-Derived Carbon Nanothreads. *Nature materials* **2015**, *14*, 43-47.

10. Zhan, H.; Zhang, G.; Tan, V. B.; Cheng, Y.; Bell, J. M.; Zhang, Y.-W.; Gu, Y., From Brittle to Ductile: A Structure Dependent Ductility of Diamond Nanothread. *Nanoscale* **2016**, *8*, 11177-11184.

11. Tabandeh-Khorshid, M.; Omrani, E.; Menezes, P. L.; Rohatgi, P. K., Tribological Performance of Self-Lubricating Aluminum Matrix Nanocomposites: Role of Graphene Nanoplatelets. *Engineering Science and Technology, an International Journal* **2016**, *19*, 463-469.

12. Moghadam, A. D.; Omrani, E.; Menezes, P. L.; Rohatgi, P. K., Mechanical and Tribological Properties of Self-Lubricating Metal Matrix Nanocomposites Reinforced by Carbon Nanotubes (CNTs) and Graphene–a Review. *Composites Part B: Engineering* **2015**, *77*, 402-420.

13. Cranford, S. W., When Is 6 Less Than 5? Penta- to Hexa-Graphene Transition. *Carbon* **2016**, *96*, 421-428.

14. Chenoweth, K.; Van Duin, A. C.; Goddard, W. A., Reaxff Reactive Force Field for Molecular Dynamics Simulations of Hydrocarbon Oxidation. *The Journal of Physical Chemistry A* **2008**, *112*, 1040-1053.

15. Yoon, K.; Ostadhossein, A.; van Duin, A. C., Atomistic-Scale Simulations of the Chemomechanical Behavior of Graphene under Nanoprojectile Impact. *Carbon* **2016**, *99*, 58-64.

16. Cranford, S. W.; Buehler, M. J., Mechanical Properties of Graphyne. *Carbon* **2011**, *49*, 4111-4121.

17. Roman, R. E.; Kwan, K.; Cranford, S. W., Mechanical Properties and Defect Sensitivity of Diamond Nanothreads. *Nano letters* **2015**, *15*, 1585-1590.

18. Hoover, W. G., Canonical Dynamics: Equilibrium Phase-Space Distributions. *Physical review A* **1985**, *31*, 1695.

19. Nosé, S., A Unified Formulation of the Constant Temperature Molecular Dynamics Methods. *The Journal of chemical physics* **1984**, *81*, 511-519.

20. Plimpton, S., Fast Parallel Algorithms for Short-Range Molecular Dynamics. *Journal of computational physics* **1995**, *117*, 1-19.

21. Diao, J.; Gall, K.; Dunn, M. L., Atomistic Simulation of the Structure and Elastic Properties of Gold Nanowires. *Journal of the Mechanics and Physics of Solids* **2004**, *52*, 1935-1962.

22. Jang, D.; Li, X.; Gao, H.; Greer, J. R., Deformation Mechanisms in Nanotwinned Metal Nanopillars. *Nature nanotechnology* **2012**, *7*, 594-601.

23. Wang, J.; Lu, C.; Wang, Q.; Xiao, P.; Ke, F.; Bai, Y.; Shen, Y.; Liao, X.; Gao, H., Influence of Microstructures on Mechanical Behaviours of Sic Nanowires: A Moleculardynamics Study. *Nanotechnology* **2011**, *23*, 025703.

24. Agrawal, R.; Peng, B.; Gdoutos, E. E.; Espinosa, H. D., Elasticity Size Effects in Zno Nanowires– a Combined Experimental-Computational Approach. *Nano letters* **2008**, *8*, 3668-3674.

25. WenXing, B.; ChangChun, Z.; WanZhao, C., Simulation of Young's Modulus of Single-Walled Carbon Nanotubes by Molecular Dynamics. *Physica B: Condensed Matter* **2004**, *352*, 156-163.

TOC Graphic

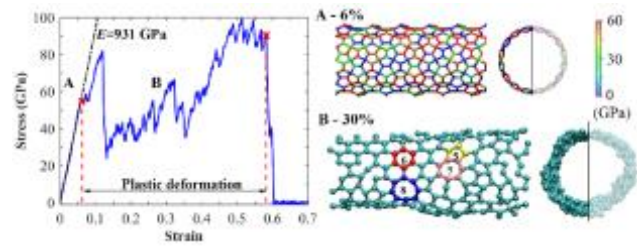